\documentclass[a4paper,onecolumn,fleqn,usenatbib]{mnras}

\usepackage[T1]{fontenc}
\usepackage{times}

\usepackage{graphicx}	
\usepackage{amsmath}	
\usepackage{amssymb}	
\usepackage{bm}

\newcommand{\md}{\mathrm{d}}
\newcommand{\me}{\mathrm{e}}
\newcommand{\mi}{\mathrm{i}}

\def\p{\partial}

\newcommand{\bJ}{\mathbf{J}}
\newcommand{\bk}{\mathbf{k}}
\newcommand{\bOm}{\mathbf{\Omega}}

\newcommand{\nn}{\nonumber}

 \title[Kinetic equation for stellar systems]{A simple, heuristic derivation of the Balescu-Lenard kinetic equation for stellar systems}

\author[C. Hamilton]{
  Chris Hamilton\thanks{E-mail: ch783@cam.ac.uk}\\
Department of Applied Mathematics and Theoretical Physics, University of
Cambridge, Wilberforce Road, Cambridge CB3 0WA, UK}

\pubyear{2020}

\begin{document}
\label{firstpage}
\pagerange{\pageref{firstpage}--\pageref{lastpage}}
\maketitle


\begin{abstract}
The unshielded nature of gravity means that stellar systems are inherently inhomogeneous.  As a result, stars do not move in straight lines. This obvious fact severely complicates the kinetic theory of stellar systems because position and velocity turn out to be poor coordinates with which to describe stellar orbits --- instead, one must use angle-action variables.  Moreover, the slow relaxation of star clusters and galaxies can be enhanced or suppressed by collective interactions (`polarisation' effects) involving many stars simultaneously. These collective effects are also present in plasmas; in that case they are accounted for by the Balescu-Lenard (BL) equation, which is a kinetic equation in velocity space.  Recently, several authors have shown how to account for both inhomogeneity and collective effects in the kinetic theory of stellar systems by deriving an angle-action generalisation of the BL equation. Unfortunately their derivations are long and complicated, involving multiple coordinate transforms, contour integrals in the complex plane, and so on.
On the other hand, Rostoker's superposition principle allows one to pretend that a long-range interacting $N$-body system, such as a plasma or star cluster, consists merely of uncorrelated particles that are `dressed' by polarisation clouds.
In this paper we use Rostoker's principle
to provide a simple, intuitive derivation of the BL equation for stellar systems which is much shorter than others in the literature.  It also allows us to straightforwardly connect the BL picture of self-gravitating kinetics to the classical `two-body relaxation' theory of uncorrelated flybys pioneered by Chandrasekhar.

\end{abstract}





\section{Introduction}

The kinetic theory of self-gravitating systems such as star clusters and galaxies is greatly complicated by the fact that unperturbed stellar orbits are not straight lines.  This is ultimately a consequence of the unshielded nature of gravity, which forces stellar systems, unlike electrostatic plasmas,  to be inherently inhomogeneous.
As a result, position and velocity turn out to be unsuitable coordinates for describing stars' motion.  Instead, assuming the unperturbed motion to be integrable, it is natural to describe orbits using angle-action variables
 \citep{Binney2008}.  Furthermore, the  long-range nature of the Newtonian gravitational force means that particles can behave \textit{collectively}. For instance, the
 internal discreteness noise of a spherical stellar cluster (simply due to the finite number of stars) gives rise to stochastic fluctuations in the cluster's internal gravitational potential. These fluctuations tend to be amplified by the collective self-gravitating motion of the stars, particularly on system-size scales \citep{Weinberg1998,Hamilton2018,Lau2019}.  Indeed, spherical clusters can support system-scale collective oscillations that, although transient, last much longer than a typical orbital period (see e.g. \citealt{Mathur1990, Weinberg1994, Heggie2020}). 
 Analogously, Debye shielding in a plasma results from the simultaneous collective behaviour of many particles via the long-range Coulomb force \citep{Ichimaru1973,Escande2018}.  
 Unfortunately, the necessity of (i) introducing angle-action variables to account for inhomogeneity and (ii) accounting for collective effects makes it difficult to develop a self-consistent kinetic theory of self-gravitating systems.  
 
\citet{Chandrasekhar1942,Chandrasekhar1943a,Chandrasekhar1943b} bypassed these difficulties by considering the evolution of a `test star' as it flies in a straight line through a fixed, homogeneous background of `field stars'. 
In his picture, the test star receives a weak impulsive kick from each field star that it encounters with some impact parameter $b$.  Chandrasekhar then pretends that the evolution of the test star's orbit consists of nothing more than an uncorrelated series of these small, discrete nudges. Subsequently, the entire stellar system can be thought of as a superposition of test stars, and the system's slow relaxation\footnote{By `relaxation' we mean here the slow evolution of the system's smooth phase space distribution function on timescales much longer than a typical orbital period. Since there is no maximum entropy state for stellar systems, we can not think of relaxation towards e.g. some universal Maxwellian distribution (\citealt{Binney2008}).} as resulting from the many impulsive two-body encounters that they experience. By integrating over all $b$ with suitably chosen cutoffs at some $b_\mathrm{min}$ and $b_\mathrm{max}$ to avoid divergences, one derives drift and diffusion coefficients that can be inserted in to a Fokker-Planck equation describing the evolution of the system in velocity or energy space (e.g. \S7.4 of \citealt{Binney2008}).  The drift and diffusion coefficients both feature the Coulomb logarithm, $\ln (b_\mathrm{max}/b_\mathrm{min})$, whose value must be prescribed by hand or calibrated from simulations \citep{Giersz1994}. Despite the crudeness of his model, Chandrasekhar's theory\footnote{In fact, this two-body relaxation picture had already been developed for electrostatic plasmas by \citet{Landau1936}, while the now-standard method for calculating drift and diffusion coefficients was formulated by \citet{Rosenbluth1957}. For a detailed history of the various kinetic theories mentioned in this paper, see \citet{Chavanis2013a,Chavanis2013b} and \citet{Heyvaerts2017}.} has proved very useful for describing collisional stellar systems like globular clusters (e.g. \citealt{Giersz1994,Vasiliev2015}).  Indeed, the canonical picture of cluster relaxation has not fundamentally changed in the intervening eight decades \citep{Heggie2003}.
 
 However, it is clear that Chandrasekhar's picture is conceptually flawed. It entirely ignores both (i) the complex quasiperiodic nature of orbits in self-gravitating systems, replacing them with straight lines, and (ii) collective behaviour, by accounting only for bare, uncorrelated two-body encounters.
A more sophisticated kinetic theory is available: the Balescu-Lenard (BL) equation, first derived for homogeneous plasmas (i.e. using position-velocity coordinates) by \citet{Balescu1960} and \citet{Lenard1960}, accounts for collective effects self-consistently. Much more recently, the angle-action analogue of the BL equation has been derived by \citet{Heyvaerts2010} starting from the BBGKY hierarchy, by \citet{Chavanis2012b} from the Klimontovich equation, by \citet{Heyvaerts2017} with a generalised Fokker-Planck approach, and finally by \citet{Fouvry2018} using techniques from stochastic calculus\footnote{Actually, it seems the BL equation for stellar systems was first derived by \citet{Luciani1987}, although they did not explicitly show how to calculate the dressed collective potential. See also \citet{Hitchcock1983,Mynick1988} for angle-action treatments of BL theory in toroidal plasmas.
}.  Thus, we finally have a self-consistent kinetic theory that accounts for both effects (i) and (ii) and can therefore be applied to stellar systems like galactic disks \citep{Fouvry2015} and globular clusters \citep{Hamilton2018}. Unfortunately, while Chandrasekhar's relaxation theory is very simple, the aforemonetioned derivations of the BL equation are exceedingly complicated for beginners.
 They all involve at least two or three coordinate transforms --- typically a Fourier transform in angle, a Laplace transform in time, and an expansion of all perturbations in biorthogonal potential-density basis pairs --- as well as several rounds of contour integration, analytic continuation, and so on.
Moreover, they differ so greatly from Chandrasekhar's intuitive picture of `one star flying past another' that the connection between the two theories is often obscured. 


On the other hand, Rostoker's superposition principle \citep{Rostoker1964a,Rostoker1964b,Diamond2010} provides us with a useful conceptual bridge between these theories.  Rostoker, who considered an electrostatic plasma, realised that one can forego the usual complex calculations involving many particles interacting simultaneously
and instead simply think of the relaxation of the plasma as a superposition of uncorrelated two-particle interactions (a la Chandrasekhar), except that the interaction potential energy between pairs of particles is `dressed' by their polarisation clouds. 
Rostoker's viewpoint allows one to derive very easily and intuitively the standard (velocity space) BL equation used in plasma kinetics (see \S 2.2.3.6 of \citealt{Diamond2010}; also \citealt{Wyld1962}). It is this simplicity that we want to take advantage of in the present paper. 

Rostoker's principle can also be applied to stellar systems, as was shown by \citet{Gilbert1968,Lerche1971} and described further by \citet{Nelson1999}. In the stellar dynamical context, Rostoker's principle has been used e.g. by \cite{Weinberg1998} to calculate the fluctuation spectrum of noise in a spherical cluster, but never to derive a kinetic equation. \citet{Kaufman1971} used a similar approach to ours to construct the quasilinear theory\footnote{That is, a kinetic theory that accounts for the effect of wave modes on particles but ignores the particles' discreteness noise (\S\ref{sec:Rostoker}).} of inhomogeneous plasmas in angle-action variables. The main purpose of the present paper is to show how to derive the BL equation in angle-action variables using Rostoker's principle, thereby offering a short, physically-motivated route to the kinetic equation describing self-gravitating systems. Our derivation is somewhat heuristic because it relies on a few assumptions, including Rostoker's principle itself, that are physically plausible but only rigorously justified via the more painstaking mathematical analyses of \citet{Heyvaerts2010,Chavanis2012b,Heyvaerts2017,Fouvry2018}, etc. It can therefore be considered an accessible `shortcut' to the final equation, to be formalised later.  It also has the clear advantage of being straightforwardly connected to the Chandrasekhar picture of two-body relaxation with which most dynamicists are familiar.

In \S\ref{sec:Rostoker} we describe the physics behind Rostoker's principle. In \S\ref{sec:Derivation} we use it to derive the action-space BL equation. Finally in \S \ref{sec:Discussion} we discuss briefly the relationship between the BL and Chandrasekhar kinetic theories, and offer some thoughts on why the Chandrasekhar approach has been so successful in describing the relaxation of globular clusters despite its manifest flaws. 


\section{Relaxation of self-gravitating systems and Rostoker's principle}
\label{sec:Rostoker}

Suppose we initialise an $N$-body simulation of a globular cluster or disk
galaxy.  We draw our initial conditions (positions and velocities, or
equivalently angles and actions) at random from some underlying mechanical equilibrium one-particle distribution function (DF) $f$. Said differently, an ensemble average of a large number of such random
realisations would reproduce the equilibrium $f$. The noise in the
gravitational potential at this stage is
pure Poisson (shot) noise due to the finite number of particles. Now, when we switch on the evolution, two things
happen:
\begin{itemize}
\item The noise gets `dressed'. Correlations develop between stars on various scales in phase
space and over a broad, continuous range of frequencies.  In the plasma analogy we would think of each particle getting its own Debye `polarisation 
cloud', decreasing its effective charge. In the gravitational case, one can think of each star as accruing a global `wake' of other stars, increasing its effective mass (for real-space diagrams of wakes see \citealt{Weinberg1989}). After dressing, the correlations between stars no longer reflect pure Poisson
noise. However the system's one-particle DF is the same, i.e. it is still in the same average mechanical equilibrium governed by $f$.  Once established,
the dressed noise bubbles away at its saturated level until there is
some secular evolution (`relaxation') that changes $f$. 
\item The initial noise excites a
\textit{transient} response. Wave modes, i.e. collective oscillations of
the underlying collisionless equilibrium system described by $f$,
are spontaneously excited on many
different spatial scales, though only at a discrete set of frequencies. In a plasma these frequencies are zeros of the dielectric function $\epsilon$. Most of the modes are uninteresting because they are of small amplitude and are
damped on the orbital timescale.  However, a few modes --- such as the $\ell=1$
sloshing mode in a spherical stellar cluster --- can be of large amplitude and only very weakly damped,
so will contribute to the potential fluctuations for much longer than an
orbital time \citep{Weinberg1994,Lau2019,Heggie2020}. 
\end{itemize}
In fact, the above considerations do not only apply at $t=0$, when the noise is constructed somewhat artificially. Instead they hold whenever the system undergoes some secular evolution and has to adjust to its new equilibrium $f$. It is a central assumption of the BL kinetic theory that the saturation of the `dressing' process
--- sometimes referred to as the `thermalisation of the bath' \citep{Fouvry2018} --- and the decay of the coherent wave modes happens on a timescale much shorter than a relaxation time.  Thus, in the BL picture, to ask how a stellar system `relaxes' is to ask how $f$ evolves slowly under the influence of the steady-state spectrum of dressed discreteness noise. 

One rigorous approach to this problem is to start from the BBGKY hierarchy of equations, which successively link the one-particle DF $f$ to the two-particle DF, the two-particle DF to the three-particle DF, and so on \citep{Heyvaerts2010}.  If one closes this hierarchy using $1/N$ as a small parameter and makes a weak-coupling approximation (i.e. ignores the rare strong encounters),
then to first order in $1/N$ the resulting kinetic equation for $f$ is the BL equation. 
However, by calculating the two-particle DF for an electrostatic plasma to first order in\footnote{Strictly, in Rostoker's case $N$ is the number of particles in a Debye sphere rather than the whole system, but this does not change the argument. To higher order in $1/N$ Rostoker's picture breaks down: roughly speaking, at $\mathcal{O}(1/N^2)$ one must include cloud-cloud interactions on top of the cloud-particle interactions discussed here.} $1/N$, \citet{Rostoker1964a,Rostoker1964b} realised that polarisation effects manifest themselves only via an effective dressing of otherwise independent two-body encounters. In other words, Rostoker's principle states that the relaxation proceeds as a superposition of uncorrelated two-body kicks like in Chandrasekhar's theory, and the inclusion of collective effects just amounts to replacing the Coulomb interaction between particles with an effective dressed interaction.
Now, before blindly applying Rostoker's principle to stellar systems, one must remember that in a homogeneous electrostatic plasma there is no dominating mean force (i.e. no macroscopic electric field). Conversely, stellar systems have a dominant mean field that determines the unperturbed motion, and every star's gravitational potential contributes to this mean field. Nevertheless, \citet{Gilbert1968} (see also \citealt{Lerche1971,Gilbert1971,Nelson1999}) showed that Rostoker's principle does hold for stellar systems provided we account for the mean field motion.  This is naturally accomplished through the introduction of angle-action variables.



\section{Derivation of the angle-action Balescu-Lenard equation from Rostoker's principle}

\label{sec:Derivation}

Consider a system of composed of $N \gg 1$ stars with equal mass $m$.  We assume the mean field potential of the system, $\Phi$, is integrable.  Then there exist angle-action coordinates $(\bm{\theta}, \bJ)$ such that a particle's mean field specific Hamiltonian $H = \mathbf{v}^2/2 + \Phi(\mathbf{x})$, where $(\mathbf{x}, \mathbf{v})$ are a star's position and velocity, does not depend on $\bm{\theta}$, i.e. $H=H(\bJ,t)$. The time dependence of $H$ arises because of the slow relaxation of the system on timescales much longer than a typical orbital period; from now on we will suppress the $t$ argument, simply writing $H=H(\bJ)$. In angle-action space the star then has mean field equations of motion $\md \bm{\theta}/\md t = \partial H/\partial \mathbf{J} \equiv \bOm(\bJ)$ and $\md \mathbf{J}/\md t = \mathbf{0}$. The angle dependence of all quantities is periodic under $\theta_i \to \theta_i + 2\pi$. Furthermore, by Jeans' theorem the slowly evolving one-particle DF of stars in phase space is also a function  of the actions alone, $f = f(\mathbf{J},t)$, which we henceforth write simply as $f(\bJ)$. We normalise $f$ so that $\int \md \bm{\theta} \md \bm{J} f = \int \md \mathbf{x} \md \mathbf{v} f = Nm$; then Poisson's equation reads $\nabla^2 \Phi = 4\pi G \int \md \mathbf{v} f$. A reader unfamiliar with any of these concepts may consult \citet{Binney2008}.


The rest of this section is dedicated to deriving the BL equation for stellar systems using Rostoker's principle.  First (\S\ref{sec:Two_stars}) we consider a dressed interaction between two stars.  Second (\S\ref{sec:master_equation}) we write down the master equation that describes the evolution of $f$. Finally (\S\ref{sec:kinetic_equation}) we bring both of these ingredients together to derive the BL equation.



\subsection{Interaction of two stars via a dressed potential}
\label{sec:Two_stars}

Inspired by the discussion of \S\ref{sec:Rostoker}, let us consider a `test' star with coordinates $(\bm{\theta}, \bJ)$ and a `field' star with coordinates $(\bm{\theta}', \bJ')$. Rostoker's principle says that we can forget about all the other stars, and treat the interaction of these two stars as if they were an isolated system with specific two-body Hamiltonian (units of (velocity)$^2$):
\begin{align}
\label{eqn:two_body_Hamiltonian}
    h = H(\bJ) + H(\bJ') + U^\mathrm{d}(\bm{\theta},\bJ, \bm{\theta}',\bJ').
\end{align}
Here $U^\mathrm{d}(\bm{\theta},\bJ, \bm{\theta}',\bJ')$ is the dressed specific potential energy between a particle at phase space location $(\bm{\theta}, \bJ)$ and a particle at $(\bm{\theta}', \bJ')$. It consists of the usual Newtonian attraction plus polarisation effects; if we ignore polarisation then $U^\mathrm{d} \to -Gm/\vert \mathbf{x} - \mathbf{x}'\vert$, and we straightforwardly recover the classical two-body encounter theory detailed in Appendix L of \citet{Binney2008}. 

Let us expand $U^\mathrm{d}$ as a Fourier series in the angle variables:
\begin{align}
\label{eqn:dressed_potential_fluctuation}
U^\mathrm{d}(\bm{\theta},\mathbf{J}, \bm{\theta}',\mathbf{J}')= \sum_{\bk \bk'} \me^{\mi(\bk \cdot \bm{\theta} - \bk' \cdot \bm{\theta}') } \, U^\mathrm{d}_{\bk \bk'}(\bJ, \bJ') \equiv m \sum_{\bk \bk'} \me^{\mi(\bk \cdot \bm{\theta} - \bk' \cdot \bm{\theta}') }  \, {\psi}^\mathrm{d}_{\bk \bk'}(\bJ, \bJ', \bk' \cdot \bOm'),
\end{align}
where $\bk, \bk'$ are vectors with (dimensionless) integer entries, and we employed the notation $\bOm \equiv \bOm(\mathbf{J})$ and $\bOm' \equiv \bOm(\mathbf{J}')$.  In this equation, to connect with more rigorous BL derivations, we have introduced the quantity $\psi^\mathrm{d}_{\bk\bk'}(\bJ, \bJ', \omega)$, which is the generic form of the (frequency-dependent) dressed potential fluctuation when Fourier transformed in angles \citep{Chavanis2012b,Fouvry2018}. We will not write an explicit formula for $\psi^\mathrm{d}_{\bk\bk'}$ here. In a plasma it is simply equal to the Fourier-transformed Coulomb interaction energy divided by the frequency-dependent dielectric function $\epsilon$.  For a brief review of how to calculate $\psi^\mathrm{d}_{\bk\bk'}$ for self-gravitating systems, see Appendix G of \citet{Fouvry2018}. For a much broader discussion, as well as references to the extensive literature on the subject, see \citet{Nelson1999} and \citet{Binney2008}.

At a formal level $\psi^\mathrm{d}_{\bk\bk'}(\bJ, \bJ', \omega)$ is defined for arbitrary frequencies $\omega$. However, to be consistent with the assumptions of BL theory (\S\ref{sec:Rostoker}) $U^\mathrm{d}$ must not have any explicit time dependence. Heuristically, in a steady state the stars can only input power in to the dressed potential fluctuations at their orbital frequencies and integer multiples thereof. The mathematical details are covered in \S2 of \citet{Heyvaerts2017}; the upshot is that $\psi^\mathrm{d}_{\bk\bk'}$ only contributes to the fluctuations at frequencies\footnote{The spectrum of the noise encapsulated in $U^\mathrm{d}$ is still broad rather than discrete, because we consider fluctuations arising from field stars with all possible actions $\bJ'$, so that $\bk'\cdot \bOm'$ takes a continuous range of values provided $N$ is large enough \citep{Luciani1987}.} $\omega = \bk' \cdot \bOm'$.  Note that although we appear to have broken the symmetry between test and field stars with this choice, we will see in a moment that their interaction over long timescales washes out to zero unless they are in resonance, i.e. unless $\bk \cdot \bOm = \bk' \cdot \bOm'$, restoring the symmetry\footnote{Note also that the interaction potential $U^\mathrm{d}$ itself appears from \eqref{eqn:dressed_potential_fluctuation} to be asymmetric under interchange of test and field star coordinates. Despite this, \citet{Heyvaerts2017} showed that to sufficient accuracy in the small parameter $1/N$ we may take $\psi^{\mathrm{d}}_{\bk\bk'}(\bJ,\bJ',\omega) = \psi^{\mathrm{d}}_{\bk'\bk}(\bJ',\bJ,\omega)^*$.  Together with the identity \eqref{eqn:identity}, this means that $U^\mathrm{d}$ is in fact symmetric, justifying our use of a single two-body Hamiltonian \eqref{eqn:two_body_Hamiltonian} from which we derive the motion of both stars.}.

We expect $\vert U^\mathrm{d} /  H\vert \sim \mathcal{O}(1/N)$, so we can treat $U^\mathrm{d}$ as a perturbation. To zeroth order in this perturbation the test and field stars just follow their mean field trajectories $\bm{\theta} = \bm{\theta}_0 + \bOm t$ and $\bm{\theta}' = \bm{\theta}'_0 + \bOm' t$ indefinitely.  To first order, the result of their interaction is to nudge each other to new actions $\mathbf{J}+\delta \mathbf{J}$ and $\mathbf{J}'+\delta \mathbf{J}'$ respectively. We can calculate $\delta \bJ$ using Hamilton's equation:
\begin{align}
\delta \bJ(\bm{\theta}_0, \bJ, \bm{\theta}_0', \bJ', \tau) = - \int_{0}^\tau \md t \, \frac{\p h}{\p \bm{\theta}} 
=
-  m
 \sum_{\bk \bk'} \mi \bk \, \psi^\mathrm{d}_{\bk \bk'}(\bJ, \bJ', \bk' \cdot \bOm') \, \me^{\mi(\bk \cdot \bm{\theta}_0 - \bk' \cdot \bm{\theta}_0')}  \frac{\me^{\mi(\bk \cdot \bOm - \bk' \cdot \bOm')\tau}-1}{\mi (\bk \cdot \bOm - \bk' \cdot \bOm')} ,
 \label{eqn:delta_J} 
\end{align}
where in the second step we substituted the mean field trajectories. The result for $\delta \bJ' = - \int_0^\tau \md t \, \partial h/\partial \bm{\theta}'$ is the same as \eqref{eqn:delta_J} except we replace the first factor $\mi \bk \to -\mi \bk'$.



\subsection{The master equation}
\label{sec:master_equation}

\begin{figure}
\centering
\includegraphics[width=0.5\linewidth,trim={0cm 0.7cm 0cm 0.3cm},clip]{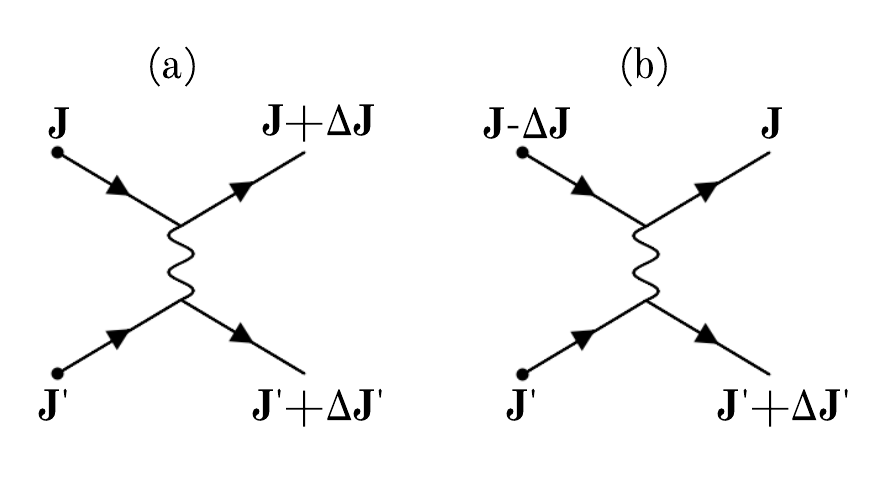}
\caption{Two possibilities for `scattering' in action coordinates. A test star's action changes by $\Delta \mathbf{J}$ during an `encounter' with a field star whose action changes from $\mathbf{J}'$ to $\bJ' + \Delta \bJ'$.
The DF $f(\bJ)$ will be decremented if the test star is kicked out of state $\bJ$ (process (a)), but will be incremented if it is kicked in to state $\bJ$ from $\bJ-\Delta \bJ$ (process (b)). To write down a master equation we must also account for the inverse processes (by reversing the directions of the arrows). In a plasma, it can be useful to think of the two particles exchanging a `plasmon' (wiggly line) with well-defined energy and momentum \citep{Pines1962,Wyld1962}. One can think of gravitational interactions similarly, but the analogy is not so useful here because we are working in action space rather than momentum (or velocity) space.}
\label{fig:BL_Feynman}
\end{figure}

Following Rostoker, we consider the relaxation of our entire system to consist of nothing more than an uncorrelated set of dressed two-body encounters.
Then it is easy to write down a master equation for the DF $f(\mathbf{J})$.  To do so, we account for (a) test stars being nudged out of the state $\bJ$ and in to some new state $\bJ+\Delta \bJ$, as illustrated in Figure \ref{fig:BL_Feynman}a, and (b) test stars being nudged in to the state $\bJ$ from $\bJ - \Delta \bJ$, as in Figure \ref{fig:BL_Feynman}b. Processes (a) and (b) are both deterministic and time-reversible,
so we must also account for their inverses, which would be represented by the same diagrams but with the arrows pointing in the opposite direction.  For the purposes of writing down the master equation we leave $\Delta \bJ$ and $\Delta \bJ'$ arbitrary; in \S\ref{sec:kinetic_equation} we will incorporate the physical values $\delta \bJ$  (equation \eqref{eqn:delta_J}) and $\delta \bJ'$ which correspond to a specific choice of initial conditions $(\bm{\theta}_0, \mathbf{J},\bm{\theta}_0',\mathbf{J}')$.

Let the \textit{transition rate density} of process (a) be $w(\Delta \mathbf{J}, \Delta \mathbf{J}' \vert \mathbf{J}, \mathbf{J}')$. This quantity is defined such that $w(\Delta \mathbf{J}, \Delta \mathbf{J}' \vert \mathbf{J}, \mathbf{J}')\,  \md \Delta \bJ \md \Delta \bJ' \tau$ is the probability that a given test star with initial action $\bJ$ is scattered to the volume of phase space $\md \Delta \bJ$ around $\bJ+\Delta \bJ$, by a given field star with action $\bJ'$ that is itself scattered to the volume element $\md \Delta \bJ'$ around $\bJ'+\Delta \bJ'$, in a time interval $\tau$ that is much longer than an orbital period but much shorter than the relaxation time (c.f. \S7.4.1 of \citealt{Binney2008}). Similarly, let the transition rate density of process (b) be $w(\Delta \mathbf{J}, \Delta \mathbf{J}' \vert \mathbf{J}-\Delta \bJ,\mathbf{J}')$.
We assume the system's equilibrium state is invariant under time reversal\footnote{See \S4.1 of \citealt{Nelson1999}; this necessarily excludes, for instance, rotating galaxies.}.  Then each process has the same transition rate as its inverse.  Finally, we consider perturbations not from a single field star but an ensemble of field stars, noting that the average number of field stars in the action space volume element surrounding $\bJ'$ is $(2\pi)^3 f(\bJ') \,\md \bJ' / m$. Integrating over all field star actions $\bJ'$ and all possible kicks $\Delta \mathbf{J}, \Delta \mathbf{J}'$, the master equation reads\footnote{c.f. \S\S1-2 of \citet{Wyld1962}.  The factor $1/2$ in the integrand arises because, for instance, the term $(2\pi)^{3}m^{-1}\int\md \bJ' \md \Delta \bJ \md \Delta \bJ' w(\Delta \bJ \Delta \bJ' \vert \bJ,\bJ')f(\bJ)f(\bJ') $ arising from diagram (a) already accounts for all the ways in which the test star can be kicked out of the state $\bJ$.  Thus we must take care not to over-count when writing down the contribution of diagram (b).}
\begin{align}
 \frac{\p f(\mathbf{J})}{\p t} 
 = \frac{(2\pi)^3}{m} \int \md \bJ' \md \Delta \bJ \md \Delta \bJ' 
\times \frac{1}{2} \Big[ &w(\Delta \mathbf{J}, \Delta \mathbf{J}' \vert  \mathbf{J}, \mathbf{J}') [ -f(\bJ) f(\bJ') + f(\bJ +\Delta \bJ)f(\bJ' + \Delta \bJ')] \nn
 \\
 + &w(\Delta \mathbf{J}, \Delta \mathbf{J}' \vert  \mathbf{J}-\Delta \bJ,\mathbf{J}')
[ f(\bJ-\Delta \bJ) f(\bJ') - f(\bJ)f(\bJ' + \Delta \bJ') ] 
 \Big].
\end{align}
Now we expand the integrand for weak interactions, i.e. for $\Delta \bJ, \Delta \bJ' \ll \bJ, \bJ'$.  In particular this means we are throwing away close two-body encounters --- in Chandrasekhar's theory this is achieved by cutting off the integral over impact parameter at $b_\mathrm{min}$ (see \citealt{Binney2008}, \S7.4.2a).  As a result, we find that terms to zeroth and first order in the small quantities all cancel. Keeping only the second order terms we get 
\begin{align}
\label{eqn:kinetic_equation_A_B}
 \frac{\p f(\mathbf{J})}{\p t} 
 = 
 \frac{\p}{\p \bJ} \cdot \int \md \bJ' 
\left[ \mathsf{A}\cdot  f(\bJ')\frac{\p f}{\p \bJ} +
\mathsf{B} \cdot f(\bJ)\frac{\p f}{\p \bJ'}
\right].
\end{align}
Here $\mathsf{A}$ is a $3\times 3$ matrix that depends on $\bJ$ and $\bJ'$:
\begin{align}
\label{eqn:A}
\mathsf{A}(\bJ,\bJ') &= \frac{(2\pi)^3}{2m} \int \md \Delta \bJ \md \Delta \bJ' w(\Delta \mathbf{J}, \Delta \mathbf{J}' \vert \mathbf{J}, \mathbf{J}') \, \Delta \bJ \Delta \bJ \equiv \frac{(2\pi)^3}{2m} \frac{\langle \Delta \bJ \Delta \bJ \rangle_\tau}{\tau},
\end{align}
where $\langle \Delta \bJ \Delta \bJ \rangle_\tau$ is the expectation value of $\Delta \bJ \Delta \bJ$ after a time interval $\tau$ for a given test star action $\bJ$ and field star action $\bJ'$. The matrix $\mathsf{B}$ is identical to $\mathsf{A}$ except we replace $\langle \Delta \bJ \Delta \bJ \rangle_\tau \to \langle \Delta \bJ \Delta \bJ' \rangle_\tau$. In a homogeneous system, $\mathbf{J}$ and $\mathbf{J}'$ can be taken to be the (suitably normalised) linear momenta of the test and field star; then momentum conservation implies $\Delta \bJ = - \Delta \bJ'$ so that $\mathsf{A} = -\mathsf{B}$ and one easily recovers equation (2.36) of \citet{Diamond2010}.


\subsection{The kinetic equation}
\label{sec:kinetic_equation}

Now we put the results of \S\ref{sec:Two_stars} and \S\ref{sec:master_equation} together. Since by Jeans' theorem stars are uniformly distributed in the angle variables, we can calculate the expectation value $\langle \Delta \bJ \Delta \bJ \rangle_\tau$ by averaging over initial phases $\bm{\theta}_0, \bm{\theta}_0'$. Thus we have
\begin{align}
\label{eqn:expectation_value}
    \langle \Delta \bJ \Delta \bJ \rangle_\tau = \int \frac{\md \bm{\theta}_0}{(2\pi)^3} \frac{\md \bm{\theta}_0'}{(2\pi)^3} \,\delta \bJ(\bm{\theta}_0, \bJ, \bm{\theta}_0', \bJ', \tau) \, \delta \bJ(\bm{\theta}_0, \bJ, \bm{\theta}_0', \bJ', \tau),
\end{align}
where $\delta \bJ$ is given in equation \eqref{eqn:delta_J}. Our approach here is completely analogous to the Chandrasekhar method in which one calculates expectation values of velocity perturbations by integrating over all impact parameters $b$ (\citealt{Binney2008}, Appendix L). 
Plugging \eqref{eqn:expectation_value} and \eqref{eqn:delta_J} in to \eqref{eqn:A} and taking the limit $\tau \to \infty$ (since we are interested in the dynamics on timescales much longer than a typical orbital period $\sim 1/\vert \bOm \vert$) we find 
\begin{align}
\mathsf{A}(\bJ, \bJ') = \frac{m}{2(2\pi)^3} \int \md \bm{\theta}_0 \md \bm{\theta}_0'
\sum_{\bk \bk'} \sum_{\mathbf{q} \mathbf{q}'} \bk \, \mathbf{q} \, & \psi^\mathrm{d}_{\bk \bk'}(\bJ, \bJ', \bk' \cdot \bOm') \, \psi^\mathrm{d}_{\mathbf{q} \mathbf{q}'}(\bJ, \bJ', \mathbf{q}' \cdot \bOm') \, \me^{\mi(\bk + \mathbf{q}) \cdot \bm{\theta}_0}
\me^{-\mi(\bk' + \mathbf{q}') \cdot \bm{\theta}'_0} \nn \\
&\times \lim_{\tau\to\infty} \left[ \tau^{-1}\,
\frac{\me^{\mi(\bk \cdot \bOm - \bk' \cdot \bOm')\tau}-1}{\bk \cdot \bOm - \bk' \cdot \bOm'} \,
\frac{\me^{\mi(\mathbf{q} \cdot \bOm - \mathbf{q}' \cdot \bOm')\tau}-1}{\mathbf{q} \cdot \bOm - \mathbf{q}' \cdot \bOm'}\right].
\end{align}
The integrals over $\bm{\theta}_0$ and $\bm{\theta}_0'$ return $(2\pi)^3\delta_{ -\mathbf{q}}^{\bk}$ and $(2\pi)^3\delta_{-\mathbf{q}'}^{\bk'}$ respectively. We also make use the following two identities (see equations (44)-(45) of \citealt{Heyvaerts2017}):
\begin{align}
\label{eqn:identity}
    \psi^\mathrm{d}_{\bk \bk'}(\bJ,\bJ',\omega) = [\psi^\mathrm{d}_{-\bk, -\bk'}(\bJ,\bJ',-\omega)]^*, \,\,\,\,\,\,\,\,\,\,\mathrm{and}\,\,\,\,\,\,\,\,\,\, \lim_{\tau \to \infty} \left[ \vert \me^{\mi x\tau}-1 \vert^2/x^2\tau\right] = 2\pi\delta (x),
\end{align}
the first of which follows from the fact that $U^\mathrm{d}$ is real. The result is
\begin{align}
\label{eqn:A_final}
\mathsf{A}(\bJ, \bJ') = \pi (2\pi)^3 m \sum_{\bk \bk'}
\bk \,\bk \,
\delta (\bk \cdot \bOm - \bk' \cdot \bOm') \vert \psi^\mathrm{d}_{\bk \bk'}(\bJ, \bJ', \bk' \cdot \bOm') \vert^2 .
\end{align}
The delta function in this equation explicitly enforces the resonance condition $\bk \cdot \bOm = \bk' \cdot \bOm'$ mentioned in \S\ref{sec:Two_stars}. The calculation of $\mathsf{B}$ is very similar; the result is identical to \eqref{eqn:A_final} 
except we replace the factor $\bk \bk$ with $- \bk \bk'$. 
Putting the explicit formulae for $\mathsf{A}(\bJ,\bJ')$ and $\mathsf{B}(\bJ,\bJ')$ in to the kinetic equation \eqref{eqn:kinetic_equation_A_B} we arrive at the action-space Balescu-Lenard equation (e.g. equation (32) of \citealt{Fouvry2018}):
\begin{align}
\frac{\p f(\bJ)}{\p t} = \pi (2\pi)^3 m \frac{\p}{\p \bJ} \cdot  \sum_{\bk \bk'} \bk \int \md \bJ' \delta (\bk \cdot \bOm - \bk' \cdot \bOm') \vert \psi^\mathrm{d}_{\bk \bk'}(\bJ, \bJ', \bk' \cdot \bOm') \vert^2
\left( \bk \cdot \frac{\p}{\p \bJ}
-
\bk' \cdot \frac{\p}{\p \bJ'}
\right) f(\bJ) f(\bJ').
\label{eqn:BL}
\end{align}


\section{Discussion and conclusions}

\label{sec:Discussion}

In this paper we have used Rostoker's superposition principle to provide a heuristic, accessible derivation of the Balescu-Lenard kinetic equation \eqref{eqn:BL} for self-gravitating systems.  This equation accounts for both the inherent inhomogeneity of stellar systems and for collective interactions. The derivation we have given is much simpler than others in the literature, being strongly analogous to the familiar Chandrasekhar picture of two-body relaxation: stars suffer a series of uncorrelated, pairwise kicks, and the integration over initial phases $\bm{\theta}_0$, $\bm{\theta}_0'$ is analogous to the classical integral over impact parameter $b$.  Another advantage of the present formalism is that by keeping track of both $\Delta \bJ$ and $\Delta \bJ'$ in the master equation we only needed to calculate the changes in action to first order in perturbation theory (equation \eqref{eqn:delta_J}).  In the traditional Fokker-Planck approach to deriving kinetic equations in angle-action variables one only keeps track of $\Delta \bJ$; then to calculate the `frictional' part of the equation (essentially our $\mathsf{B}$) one must go to second order, complicating the derivation significantly \citep{Binney1988,Heyvaerts2017}. The key disadvantages of our approach are a lack of formal mathematical rigour and the need to quote a few properties of the dressed interaction energy (e.g. equation \eqref{eqn:identity}) that we did not derive from scratch.

If one drops collective effects in our derivation, then instead of the BL equation one arrives at the `inhomogeneous Landau equation': this is the same result as \eqref{eqn:BL} but with $\psi^\mathrm{d} \to -G/\vert \mathbf{x}-\mathbf{x}'\vert$, i.e. with the dressed interaction potential replaced by the bare Newtonian interaction. If one further ignores the inhomogeneity of the system by treating encounters as `local', then one can replace stellar orbits by straight lines and Chandrasekhar's picture can be recovered explicitly. For further discussion of the relation between the BL equation, the Landau equation, and the classical two-body relaxation theory, we refer the reader to \citet{Chavanis2012a,Chavanis2013b}.

Rostoker's principle perhaps gives some clues as to why Chandrasekhar's theory of two-body relaxation has worked so effectively in describing the evolution of globular clusters despite its ignorance of collective effects and inhomogeneity. Rostoker demonstrated that, regardless of the great complexities of the collective $N$-body problem, to first order in $1/N$ the relaxation essentially proceeds as a superposition of uncorrelated two-body kicks, just like Chandrasekhar assumed. Since $N \sim 10^6$ in globular clusters, this effective treatment is completely sufficient. As for inhomogeneity, in our description (and in reality) each pair of stars has a long-lasting, quasiperiodic interaction.  Conversely, in Chandrasekhar's theory a test and field star only disturb each other during a single flyby encounter, which for the purposes of calculation is assumed impulsive, i.e. instantaneous in time and local in space (see \S7.4.2b of \citealt{Binney2008}). However, this inaccuracy in Chandrasekhar's theory is partly offset by allowing the test star to interact with an ensemble of field stars on very similar orbits, equivalent to assuming periodic boundary conditions for a single test-field star pair \citep{Chavanis2013b}.  Overall, we see that Chandrasekhar's theory will appear to work as long as the uncertainty in the Coulomb logarithm overwhelms the errors induced by (i) using the bare interaction energy rather than the dressed one, and (ii) invoking a train of particles on straight line orbits rather than allowing for a sustained, quasiperiodic interaction between each pair of stars on mean field orbits. \citet{Hamilton2018} found that the action space flux from the largest scale self-gravitating fluctuations in a globular cluster, computed using the BL equation \eqref{eqn:BL}, was comparable to the bare flux from two-body encounters computed using Chandrasekhar's theory.
If this is true then, roughly speaking, the `error' in Chandrasekhar's prediction can always be covered up by tweaking the Coulomb logarithm by a factor of $\sim 2$.

Finally, in this paper we have utilised concepts from plasma physics developed more than half a century ago: most obviously the superposition principle of \citet{Rostoker1964a,Rostoker1964b}, but also the diagrammatic (`quasiparticle') approach pioneered by \citet{Pines1962} and \citet{Wyld1962}. In doing so we have managed to streamline an otherwise difficult and cumbersome calculation.
Nevertheless, we have only scraped the surface of the enormous literature on plasma kinetics that extends far beyond BL theory. While various tools from plasma physics have been successfully adapted to stellar systems over the years \citep{Nelson1999,Binney2008,Chavanis2013b} there are likely still many other concepts yet to be `translated' in this way. These could be useful not only for the purposes of easing calculation but also for enhancing physical understanding.



\section*{Acknowledgements}

I am very grateful to Prof. D. Heggie for a highly perceptive review and for correcting the original equation (4). I also thank J.-B. Fouvry, J. Binney, T. Heinemann, J. Magorrian, G. Wagner and M. Abazorius for many helpful comments and discussions, and to A. Ellis for proofreading the manuscript. Finally I thank A. Schekochihin, whose 2020 Zoom lectures introduced me to the diagrammatic/quasiparticle approach to kinetics \citep{Schekochihin2017}. My research is funded by a Science and Technology Facilities Council (STFC) studentship.

\section*{Data availability}
No new data were generated or analysed in support of this research.




\bibliographystyle{mnras}
\bibliography{Bibliography} 

\begin{thebibliography}{}
\makeatletter
\relax
\def\mn@urlcharsother{\let\do\@makeother \do\$\do\&\do\#\do\^\do\_\do\%\do\~}
\def\mn@doi{\begingroup\mn@urlcharsother \@ifnextchar [ {\mn@doi@}
  {\mn@doi@[]}}
\def\mn@doi@[#1]#2{\def\@tempa{#1}\ifx\@tempa\@empty \href
  {http://dx.doi.org/#2} {doi:#2}\else \href {http://dx.doi.org/#2} {#1}\fi
  \endgroup}
\def\mn@eprint#1#2{\mn@eprint@#1:#2::\@nil}
\def\mn@eprint@arXiv#1{\href {http://arxiv.org/abs/#1} {{\tt arXiv:#1}}}
\def\mn@eprint@dblp#1{\href {http://dblp.uni-trier.de/rec/bibtex/#1.xml}
  {dblp:#1}}
\def\mn@eprint@#1:#2:#3:#4\@nil{\def\@tempa {#1}\def\@tempb {#2}\def\@tempc
  {#3}\ifx \@tempc \@empty \let \@tempc \@tempb \let \@tempb \@tempa \fi \ifx
  \@tempb \@empty \def\@tempb {arXiv}\fi \@ifundefined
  {mn@eprint@\@tempb}{\@tempb:\@tempc}{\expandafter \expandafter \csname
  mn@eprint@\@tempb\endcsname \expandafter{\@tempc}}}

\bibitem[\protect\citeauthoryear{Balescu}{Balescu}{1960}]{Balescu1960}
Balescu R.,  1960, The Physics of Fluids, 3, 52

\bibitem[\protect\citeauthoryear{Binney \& Lacey}{Binney \&
  Lacey}{1988}]{Binney1988}
Binney J.,  Lacey C.,  1988, Monthly Notices of the Royal Astronomical Society,
  230, 597

\bibitem[\protect\citeauthoryear{{Binney} \& {Tremaine}}{{Binney} \&
  {Tremaine}}{2008}]{Binney2008}
{Binney} J.,  {Tremaine} S.,  2008, {Galactic Dynamics: Second Edition}.
Princeton University Press

\bibitem[\protect\citeauthoryear{Chandrasekhar}{Chandrasekhar}{1942}]{Chandrasekhar1942}
Chandrasekhar S.,  1942, Principles of Stellar Dynamics

\bibitem[\protect\citeauthoryear{Chandrasekhar}{Chandrasekhar}{1943a}]{Chandrasekhar1943a}
Chandrasekhar S.,  1943a, The Astrophysical Journal, 97, 255

\bibitem[\protect\citeauthoryear{Chandrasekhar}{Chandrasekhar}{1943b}]{Chandrasekhar1943b}
Chandrasekhar S.,  1943b, The Astrophysical Journal, 97, 263

\bibitem[\protect\citeauthoryear{Chavanis}{Chavanis}{2012a}]{Chavanis2012a}
Chavanis P.-H.,  2012a, The European Physical Journal Plus, 127, 19

\bibitem[\protect\citeauthoryear{Chavanis}{Chavanis}{2012b}]{Chavanis2012b}
Chavanis P.-H.,  2012b, Physica A: Statistical Mechanics and its Applications,
  391, 3680

\bibitem[\protect\citeauthoryear{Chavanis}{Chavanis}{2013a}]{Chavanis2013a}
Chavanis P.-H.,  2013a, The European Physical Journal Plus, 128, 126

\bibitem[\protect\citeauthoryear{Chavanis}{Chavanis}{2013b}]{Chavanis2013b}
Chavanis P.-H.,  2013b, Astronomy \& Astrophysics, 556, A93

\bibitem[\protect\citeauthoryear{Diamond, Itoh  \& Itoh}{Diamond
  et~al.}{2010}]{Diamond2010}
Diamond P.~H.,  Itoh S.-I.,   Itoh K.,  2010, Modern Plasma Physics: Volume 1,
  Physical Kinetics of Turbulent Plasmas.
Cambridge University Press

\bibitem[\protect\citeauthoryear{Escande, B{\'e}nisti, Elskens, Zarzoso  \&
  Doveil}{Escande et~al.}{2018}]{Escande2018}
Escande D.,  B{\'e}nisti D.,  Elskens Y.,  Zarzoso D.,   Doveil F.,  2018,
  Reviews of Modern Plasma Physics, 2, 9

\bibitem[\protect\citeauthoryear{Fouvry \& Bar-Or}{Fouvry \&
  Bar-Or}{2018}]{Fouvry2018}
Fouvry J.-B.,  Bar-Or B.,  2018, Monthly Notices of the Royal Astronomical
  Society, 481, 4566

\bibitem[\protect\citeauthoryear{Fouvry, Pichon, Magorrian  \& Chavanis}{Fouvry
  et~al.}{2015}]{Fouvry2015}
Fouvry J.-B.,  Pichon C.,  Magorrian J.,   Chavanis P.-H.,  2015, Astronomy \&
  Astrophysics, 584, A129

\bibitem[\protect\citeauthoryear{Giersz \& Heggie}{Giersz \&
  Heggie}{1994}]{Giersz1994}
Giersz M.,  Heggie D.~C.,  1994, Monthly Notices of the Royal Astronomical
  Society, 268, 257

\bibitem[\protect\citeauthoryear{Gilbert}{Gilbert}{1968}]{Gilbert1968}
Gilbert I.~H.,  1968, The Astrophysical Journal, 152, 1043

\bibitem[\protect\citeauthoryear{Gilbert}{Gilbert}{1971}]{Gilbert1971}
Gilbert I.~H.,  1971, in International Astronomical Union Colloquium. pp 3--10

\bibitem[\protect\citeauthoryear{Hamilton, Fouvry, Binney  \& Pichon}{Hamilton
  et~al.}{2018}]{Hamilton2018}
Hamilton C.,  Fouvry J.-B.,  Binney J.,   Pichon C.,  2018, Monthly Notices of
  the Royal Astronomical Society, 481, 2041

\bibitem[\protect\citeauthoryear{Heggie \& Hut}{Heggie \&
  Hut}{2003}]{Heggie2003}
Heggie D.,  Hut P.,  2003, The Gravitational Million–Body Problem: A
  Multidisciplinary Approach to Star Cluster Dynamics.
Cambridge University Press

\bibitem[\protect\citeauthoryear{Heggie, Breen  \& Varri}{Heggie
  et~al.}{2020}]{Heggie2020}
Heggie D.~C.,  Breen P.~G.,   Varri A.~L.,  2020, Monthly Notices of the Royal
  Astronomical Society, 492, 6019

\bibitem[\protect\citeauthoryear{Heyvaerts}{Heyvaerts}{2010}]{Heyvaerts2010}
Heyvaerts J.,  2010, Monthly Notices of the Royal Astronomical Society, 407,
  355

\bibitem[\protect\citeauthoryear{Heyvaerts, Fouvry, Chavanis  \&
  Pichon}{Heyvaerts et~al.}{2017}]{Heyvaerts2017}
Heyvaerts J.,  Fouvry J.-B.,  Chavanis P.-H.,   Pichon C.,  2017, Monthly
  Notices of the Royal Astronomical Society, 469, 4193

\bibitem[\protect\citeauthoryear{Hitchcock, Hazeltine  \& Mahajan}{Hitchcock
  et~al.}{1983}]{Hitchcock1983}
Hitchcock D.,  Hazeltine R.,   Mahajan S.,  1983, The Physics of fluids, 26,
  2603

\bibitem[\protect\citeauthoryear{Ichimaru}{Ichimaru}{1973}]{Ichimaru1973}
Ichimaru S.,  1973, Basic principles of plasma physics: a statistical approach.
CRC Press

\bibitem[\protect\citeauthoryear{Kaufman}{Kaufman}{1971}]{Kaufman1971}
Kaufman A.~N.,  1971, The Physics of Fluids, 14, 387

\bibitem[\protect\citeauthoryear{Landau}{Landau}{1936}]{Landau1936}
Landau L.,  1936, Phys. Z. Sowjetunion, 10, 154

\bibitem[\protect\citeauthoryear{Lau \& Binney}{Lau \& Binney}{2019}]{Lau2019}
Lau J.~Y.,  Binney J.,  2019, Monthly Notices of the Royal Astronomical
  Society, 490, 478

\bibitem[\protect\citeauthoryear{Lenard}{Lenard}{1960}]{Lenard1960}
Lenard A.,  1960, Annals of Physics, 10, 390

\bibitem[\protect\citeauthoryear{Lerche}{Lerche}{1971}]{Lerche1971}
Lerche I.,  1971, The Astrophysical Journal, 166, 207

\bibitem[\protect\citeauthoryear{Luciani \& Pellat}{Luciani \&
  Pellat}{1987}]{Luciani1987}
Luciani J.,  Pellat R.,  1987, Journal de Physique, 48, 591

\bibitem[\protect\citeauthoryear{Mathur}{Mathur}{1990}]{Mathur1990}
Mathur S.~D.,  1990, Monthly Notices of the Royal Astronomical Society, 243,
  529

\bibitem[\protect\citeauthoryear{Mynick}{Mynick}{1988}]{Mynick1988}
Mynick H.~E.,  1988, Journal of plasma physics, 39, 303

\bibitem[\protect\citeauthoryear{Nelson \& Tremaine}{Nelson \&
  Tremaine}{1999}]{Nelson1999}
Nelson R.~W.,  Tremaine S.,  1999, Monthly Notices of the Royal Astronomical
  Society, 306, 1

\bibitem[\protect\citeauthoryear{Pines \& Schrieffer}{Pines \&
  Schrieffer}{1962}]{Pines1962}
Pines D.,  Schrieffer J.~R.,  1962, Physical Review, 125, 804

\bibitem[\protect\citeauthoryear{Rosenbluth, MacDonald  \& Judd}{Rosenbluth
  et~al.}{1957}]{Rosenbluth1957}
Rosenbluth M.~N.,  MacDonald W.~M.,   Judd D.~L.,  1957, Physical Review, 107,
  1

\bibitem[\protect\citeauthoryear{Rostoker}{Rostoker}{1964a}]{Rostoker1964a}
Rostoker N.,  1964a, The Physics of Fluids, 7, 479

\bibitem[\protect\citeauthoryear{Rostoker}{Rostoker}{1964b}]{Rostoker1964b}
Rostoker N.,  1964b, The Physics of Fluids, 7, 491

\bibitem[\protect\citeauthoryear{Schekochihin}{Schekochihin}{2017}]{Schekochihin2017}
Schekochihin A.~A.,  2017, Lecture notes on kinetic theory and
  magnetohydrodynamics of plasmas

\bibitem[\protect\citeauthoryear{Vasiliev}{Vasiliev}{2015}]{Vasiliev2015}
Vasiliev E.,  2015, Monthly Notices of the Royal Astronomical Society, 446,
  3150

\bibitem[\protect\citeauthoryear{Weinberg}{Weinberg}{1989}]{Weinberg1989}
Weinberg M.~D.,  1989, Monthly Notices of the Royal Astronomical Society, 239,
  549

\bibitem[\protect\citeauthoryear{{Weinberg}}{{Weinberg}}{1994}]{Weinberg1994}
{Weinberg} M.~D.,  1994, The Astrophysical Journal, 421, 481

\bibitem[\protect\citeauthoryear{Weinberg}{Weinberg}{1998}]{Weinberg1998}
Weinberg M.~D.,  1998, Monthly Notices of the Royal Astronomical Society, 297,
  101

\bibitem[\protect\citeauthoryear{Wyld \& Pines}{Wyld \& Pines}{1962}]{Wyld1962}
Wyld H.,  Pines D.,  1962, Physical Review, 127, 1851

\makeatother
\end{thebibliography}


\bsp	
\label{lastpage}

\end{document}